\documentclass[12pt,preprint,flushrt]{aastex}

\begin{document}
\newcommand{\etal}{{\it et al.}}
\include{idl}

\title{CHANDRA Detection of the AM CVn binary ES Cet (KUV 01584-0939)}
\author{Tod E. Strohmayer}
\affil{Laboratory for High Energy Astrophysics, NASA's Goddard Space Flight 
Center, Greenbelt, MD 20771; stroh@clarence.gsfc.nasa.gov}

\begin{abstract}

We report on {\it Chandra} ACIS observations of the ultracompact AM
CVn binary ES Cet. This object has a 10.3 minute binary period and is
the most compact of the confirmed AM CVn systems. We have, for the
first time, unambiguously detected the X-ray counterpart to ES Cet. In
a 20 ksec ACIS-S image a point-like X-ray source is found within 1''
of the catalogued optical position. The mean countrate in ACIS-S is
0.013 s$^{-1}$, and there is no strong evidence for variability.  We
folded the X-ray data using the optical ephemeris of Warner \& Woudt,
but did not detect any significant modulation. If an $\approx 100 \%$
modulation similar to those seen in the ultracompact candidates V407
Vul and RX J0806.3+1527 were present then we would have detected
it. The upper limit (3$\sigma$) to any modulation at the putative
orbital period is $\approx 15\%$ (rms). We extract the first X-ray
spectrum from ES Cet, and find that it is not well described by simple
continuum models. We find suggestive evidence for discrete spectral
components at $\approx 470$ and $890$ eV, that can be modelled as
gaussian emission lines. In comparison with recent X-ray detections of
nitrogen and neon in another AM CVn system (GP Com), it appears
possible that these features may represent emission lines from these
same elements; however, deeper spectroscopy will be required to
confirm this. Our best spectral model includes a black body continuum
with $kT = 0.8$ keV along with the gaussian lines. The unabsorbed 0.2
- 5 keV X-ray flux was $\approx 7 \times 10^{-14}$ ergs cm$^{-2}$
s$^{-1}$. The observed luminosity in the hard component is a small
fraction of the total expected accretion luminosity, most of which
should be radiated below 0.1 keV at the expected mass accretion rate
for this orbital period. We discuss the implications of our results
for the nature of ES Cet.

\end{abstract}

\keywords{Binaries: close - Stars: individual (KUV 01584-0939, ES Cet,
Cet3) - Stars: white dwarfs -- X-rays: binaries -- Gravitational
Waves}

\section{Introduction}

The AM CVn stars are a class of ultracompact, semidetached binaries
which are transferring helium from a degenerate (or semi-degenerate)
dwarf onto a companion white dwarf (see Warner 1995). There are
presently ten confirmed systems of this type (Woudt \& Warner 2003),
and an eleventh has recently been proposed (SDSS J1240-01; Roelofs et
al. 2004) based on optical spectroscopy. ES Cet (also known as KUV
01584-0939, and Cet3) was discovered in the Kosi survey for UV-excess
objects (Kondo, Noguchi \& Maehara 1983), and it is also listed in the
CV catolog of Downes, Webbink, \& Shara (1997). Its optical spectrum
was studied by Wegner, McMahan \& Boley (1987) and shows strong He II
emission lines. A strong blue continuum, likely due to accretion, is
also present and provides additional evidence that the system is a
compact binary.  Recently, Warner \& Woudt (2002) strengthened this
conclusion with the report of a 10.3 minute modulation of its V band
flux. The modulation is very likely the orbital period or a
``superhump'' period of an AM CVn binary.  The superhump periods in AM
CVn's are thought to be caused by disk warping and precession, and are
typically found just longward of the orbital periods (see Patterson et
al. 2002; Warner 1995). The 10.3 minute orbital period makes ES Cet
the most compact of the {\it bona fide} AM CVn stars.  Two additional
AM CVn candidates with shorter periods have recently been proposed;
V407 Vul (Cropper et al. 1998; Ramsay et al. 2000; Marsh \& Steeghs
2002; Wu et al. 2002; Strohmayer 2002; Strohmayer 2004a) and RX
J0806.4+1527 (Israel et al. 2002; Ramsay et al. 2002; Hakala et
al. 2003; Strohmayer 2003), however, it is still possible that the
observed periods in these systems are not the orbital periods of
ultracompact systems (see Cropper et al. 2003; Warner 2003; Norton,
Haswell \& Wynn 2004; Strohmayer 2004a).

If ES Cet is indeed an ultracompact binary, then gravitational
radiation can drive mass transfer at a rate as high as $1 \times
10^{-8} M_{\odot}$ yr$^{-1}$, and the object should be a bright soft
X-ray source.  Several AM CVn systems have been detected in the X-ray
band, including AM CVn, CR Boo and GP Com (van Teeseling \& Verbunt
1994; Ulla 1995; Eracleous, Halpern \& Patterson; and Strohmayer
2004b).  Indeed, recent XMM/Newton observations of GP Com have
provided, for the first time, both high signal to noise spectra as
well as detections of X-ray emission lines from an AM CVn system
(Strohmayer 2004b). These results indicate that X-ray spectral studies
can provide detailed information about the accretion process as well
as the composition of the accreted matter in such systems. It is
therefore important to expand the X-ray sample and identify additional
targets suitable for more comprehensive follow-up.  Moreover, as the
most compact of the confirmed AM CVn stars, ES Cet should be a strong
gravitational radiation source, and changes in the orbital period
should be measureable (see Woudt \& Warner 2003; Strohmayer 2004a).

The ROSAT all sky survey (RASS) faint source catalog shows a 0.0166
counts s$^{-1}$ source, 1RXS J020052.9-092435, within $\approx 10''$
of ES Cet, however, the source is too faint to provide any useful
X-ray colors, and ROSAT's positional accuracy is not sufficient to
unambiguously confirm it as the X-ray counterpart to ES Cet (Voges et
al 2000).  Here we present new {\it Chandra} observations which have
unambiguously identified the X-ray counterpart to ES Cet, further
solidifying its accreting, ultracompact credentials. In \S 2 we give a
brief overview of our {\it Chandra} observations and the imaging
analysis. We discuss our spectral and timing analyses in \S 3.  We
close in \S 4 with a discussion of the implications of our findings
for the nature of ES Cet.

\section{Data Extraction and Analysis}

{\it Chandra} observed the region around ES Cet for $\approx 20$ ksec
on October 2, 2003 (TT). Imaging was performed with ACIS-S using
TIMED-FAINT mode and the aimpoint was on the ACIS-S3 backside
illuminated chip. We used the FTOOLS version 5.3 to extract an image
centered on the optical position of ES Cet. We then employed the CIAO
tool {\it wavdetect} to search the image for point sources. Figure 1
shows the central portion of the ACIS-S3 image.  Seven new X-ray
sources are detected in this field with $> 4\sigma$ significance (the
sources are marked with 3'' radius circles), including a source at a
location consistent with the optical position of ES Cet ($\alpha =
02:00:52.23\; , \delta = -09:24:31.68$; J2000). Table 1 summarizes
some of the basic properties of all these sources. This object, which
we denote CXO 020052.2-092431.6, is additionally marked with the pair
of orthogonal lines in Figure 1. Its derived position is within 1'' of
the optical position of ES Cet. Based on the positional coincidence
and the gross X-ray properties described below we conclude that CXO
020052.2-092431.7 is the X-ray counterpart to ES Cet. The detection of
the source in X-rays provides additional confirmation that it is
indeed an accreting, ultracompact binary.

\section{Timing and Spectral Properties}

To probe the X-ray timing and spectral properties of ES Cet we first
used the CIAO tool {\it axbary} to correct the photon arrival times to
the solar system barycenter. For this we used the source coordinates
obtained from our imaging analysis above. We then extracted photons in
an $\approx 1.5''$ region centered on the source. This resulted in a
total of 249 photons for our analyses. Figure 2 shows a lightcurve,
using all source photons binned at 1,000 seconds. The mean countrate
is $1.25\times 10^{-2}$ s$^{-1}$. There is no strong evidence for
variability in the lightcurve.

To search for a modulation at the observed optical (presumably
orbital) period we folded the data using the ephemeris of Woudt \&
Warner (2003), but we did not detect any significant modulation. We
also computed a $Z_n^2$ power spectrum (see Buccheri et al. 1983;
Strohmayer 2002), which allows a search for periodic signals with
arbitrary harmonic content, $n$. We used $n = 3$ for consistency with
the pulse profiles from V407 Vul and RX J0806 which have this number
of significant harmonics. Consistent with the folding analysis, no
signal is detected in the $Z_3^2$ spectrum. We can place an upper
limit on the signal power using the $Z_3^2$ power level found in the
immediate vicinity of the optical frequency.  Expressed as an
amplitude, this gives a ($3\sigma$) upper limit of 15\% (rms).  This
limit is consistent with the value $a = 1/2 (I_{max} -
I_{min})/(I_{max} + I_{min})$, where $I_{max}$ and $I_{min}$ are the
maximum and minimum values of the folded profile, respectively. The
quantity $a$ is a convenient, frequently used measure of the amplitude
of a pulsed signal.

Interestingly, if the source had an X-ray modulation profile similar
to those seen from V407 Vul and RX J0806.4+1527, then we would have
easily been able to detect it. Our ability to detect a signal of a
given strength in these data can be quantified using the so called
signal sensitivity, which is the pulsed amplitude that, if present,
would be detected with at least a given confidence level. For these
data we have a $3\sigma$ sensitivity to a pulsed X-ray flux at the
known orbital period of $\approx 28\%$ (rms).

We extracted source and background spectral files, used the CIAO tools
to generate response matrices and modelled the spectrum within
XSPEC. The low energy response of ACIS is effectively reduced and
modified by a contaminant (Marshall et al. 2003). To account for this
we used an ancillary response file (ARF) corrected with the ACISABS
tool.  We began by fitting several simple continuum models, including,
power-law, black body, and thermal bremsstrahlung spectra. Using these
continuum models we found there is essentially no sensitivity to a
neutral hydrogen absorbing column, so we fixed $n_H = 0.02 \times
10^{22}$, which is the value deduced from the ``$n_H$'' tool available
at the HEASARC webpage (Dickey \& Lockman 1990). We found that the
best fitting thermal bremsstrahlung temperature was $> 100$ keV. Over
the bandpass of interest this spectral form looks essentially like a
power-law.  Since the two models have essentially the same goodness of
fit, for the sake of brevity we do not give further details of the
fits for the bremsstrahlung model.

Given the relatively modest number of counts it is perhaps surprising
that none of these continuum models alone does a very good job of
fitting the spectrum. See Table 2 for a summary of the spectral
fits. For each of these models the additional contribution to $\chi^2$
appears to be associated with localized emission at about 0.9 and 0.5
keV, respectively. These excesses can be adequately modelled with
gaussian emission line profiles.  Including these two additional
components we find that the best overall model is the black body
continuum (plus the gaussian lines). This fit has a minimum $\chi^2 =
13.5$ with 23 degrees of freedom. For comparison, the power-law model
including the two emission lines achieves a minimum $\chi^2 = 23.7$,
which is formally acceptable, but not as good as the fit using the
black body continuum with lines. We show the data and the best fit
with this model in Figure 3. Using our best spectral model, the
unabsorbed X-ray flux in the 0.2 - 5 keV band is $7.1 \times 10^{-14}$
ergs cm$^{-2}$ s$^{-1}$. An absorbing column of $2 \times 10^{20}$
cm$^{-2}$ reduces this number by 3\%.

The significance of the line components can be estimated by evaluating
the significance of the change in $\chi^2$ using the F-test.  However,
a difficulty with this procedure is that the continuum is not known
{\it a priori}. Using the power-law continuum and fitting each line
separately we find $\Delta\chi^2 = 5.4$ and 7.7 for the 470 and 890 eV
features, respectively, while including both lines simultaneously
gives $\Delta\chi^2 = 15.1$. These numbers correspond to F-test
probabilities of 0.148, 0.055, and 0.0135, respectively. Using the
blackbody continuum results in $\Delta\chi^2$ values of 20.8, 13.1,
and 35.7 for the 470 eV, 890 eV and simultaneous fits.  This yields
F-test probabilities of 0.0012, 0.021, and $5\times 10^{-6}$,
respectively. Because we do not know {\it a priori} the form of the
continuum, and since the lines are not compellingly significant
against the power-law continuum, we regard the present evidence for
the line features as suggestive, requiring confirmation with higher
signal to noise measurements.

We note, however, that recent high resolution X-ray spectral
measurements of the AM CVn system GP Com (Strohmayer 2004b) have
revealed strong nitrogen emission lines at $24.78 \AA = 500.4$ eV (N
VII Ly$\alpha$), and $\approx 29 \AA = 427.6$ eV (N VI). When observed
with the lower resolution of a CCD, such nitrogen emission could
produce a feature with a centroid energy consistent with the $\approx
500$ eV feature tentatively identified in ES Cet. Moreover, strong
lines of hydrogen- and helium-like neon were also detected in GP
Com. A possible origin for the $\approx 900$ eV feature suggested in
the spectrum of ES Cet could be emission from helium-like neon at
$\approx 13.5 \AA = 918.5$ eV, however, this would require very little
neon in the hydrogen-like ionization state (Ne X Ly$\alpha$ at $12.134
\AA = 1,021.9$ eV), since there is no evidence for an emission line
component in ES Cet at $\approx 1$ keV.

\section{Discussion and Implications}

The gravitational radiation torque on an ultracompact binary is a
strongly increasing function of the orbital frequency.  For an
$\approx 10$ minute system like ES Cet, the mass accretion rate should
approach $1\times 10^{-8}$ $M_{\odot}$ yr$^{-1}$.  Assuming that
matter falls from the inner Lagrange point, and that half the
gravitational potential energy is dissipated in the accretion disk,
the X-ray luminosity, $L_x$, can be expressed as (see, for example,
Nelemans et al. 2004),
\begin{equation}
L_x = \frac{1}{2}\frac{GM_1 \dot m}{R} \left ( 1 - \frac{R}{R_{L_1}} 
\right ) \;,
\end{equation}
where $M_1$, $R$, $R_{L_1}$, and $\dot m$ are the primary mass,
primary radius, the distance from the inner Lagrange point to the
center of the accretor, and the mass accretion rate,
respectively. With $\dot m = 10^{-8}$ $M_{\odot}$ yr$^{-1}$ this
expression predicts a luminosity of $\approx 3.4 \times 10^{34}$ ergs
s$^{-1}$. 

The X-ray flux from accreting, non-magnetic white dwarfs is likely
produced in a boundary layer whose properties are a strong function of
the mass accretion rate. At the high rates expected for ES Cet the
boundary layer should be optically thick, radiating with an effective
temperature in the range $3 - 5 \times 10^5$ K (see for example,
Pringle \& Savonije 1979; Narayan \& Popham 1993). The spectrum we see
with {\it Chandra} is clearly much harder than a simple black body at
$\approx 5 \times 10^5$ K ($0.043$ keV). The distance to ES Cet is not
well constrained, however, even at a distance of 1 kpc the observed
0.2 - 5 keV flux corresponds to a luminosity of $8.3 \times 10^{30}$
ergs s$^{-1}$, which is much less than that expected for $\dot m =
10^{-8}$ $M_{\odot}$ yr$^{-1}$. Clearly the hard X-ray flux observed
with {\it Chandra} represents a small fraction of the total accretion
luminosity.

If the accretion rate is indeed as high as $10^{-8}$ $M_{\odot}$
yr$^{-1}$, then the boundary layer is expected to be optically thick,
and the bulk of the accretion luminosity should be emitted below 100
eV. For example, a thermal spectrum with $T = 4\times 10^5$ K would
peak at $\approx 100$ eV, below the nominal {\it Chandra} band.
Moreover, even a modest interstellar column would strongly absorb such
a soft component. Although the uncertainty in $n_H$ makes it difficult
to be precise, we can put some constraints on the strength of such a
component. Assuming a 40 eV blackbody temperature, and taking $n_H =
0.02 \times 10^{22}$ cm$^{-2}$, we find that such a component could
have an intrinsic flux as large as $2 - 4 \times 10^{-13}$ ergs
cm$^{-2}$ s$^{-1}$ before a soft excess becomes evident in the lowest
energy channel of the {\it Chandra} spectrum. If the temperature were
as low as 30 eV then it would not be too difficult to hide an even
larger flux of $\approx 1 - 2 \times 10^{-12}$ ergs cm$^{-2}$
s$^{-1}$. We note that {\it Chandra's} low energy sensitivity is
further compromised by the ACIS contaminant. Better sensitivity at and
below 100 eV will be required to adequately address the issue of the
bolometric accretion luminosity of ES Cet.  A distance measurement
would also be extremely useful in constraining the energetics of the
system.

The detection of ES Cet as an X-ray source further solidifies its
ultracompact credentials. Comparisons with the two ultracompact
candidates; V407 Vul and RX J0806.4+1527 could therefore lead to
insights as to the true nature of these systems.  Although ES Cet has
a photometric period similar to both of the ultracompact candidates,
there would appear to be important differences between the systems. ES
Cet has a much harder spectrum than either V407 Vul or J0806, which
both show essentially no emission above 1 keV. Thus, if accretion
powers the emission in all three sources, there must be something
different about the nature of the accretion. The lack of strong
variability in ES Cet suggests the X-ray emission covers a large
azimuthal extent on the white dwarf, as would likely be produced from
a boundary layer. On the other hand the pulsations in V407 Vul and
J0806 suggest localized emission on the white dwarf surface.

The detection with {\it Chandra} of hard ($> 2$ keV) emission from ES
Cet as well as the suggestive evidence for line features indicates
that a hard, optically thin, line dominated spectral component similar
to that seen from GP Com can still be produced even at very high mass
accretion rates. Clearly deeper high resolution spectroscopy of ES Cet
as well as GP Com, with, for example, XMM/Newton, has the potential to
probe in detail the composition and physical properties of the
boundary layers in these systems. Moreover, comparisons between a low
$\dot m$ system such as GP Com, and a high $\dot m$ system like ES Cet
can provide important information on the mass accretion rate
dependence of the boundary layer spectrum and structure.

\centerline{\bf References}

\noindent{} Buccheri, R. et al. 1983, A\&A, 128, 245.

\noindent{} Cropper, M., Ramsay, G., Wu, K. \& Hakala, P. 2003, ASP
Conference Series to be published in Proc. Cape Town Workshop on
magnetic CVs, held Dec 2002, (astro-ph/0302240).

\noindent{} Cropper, M. et al. 1998, MNRAS, 293, L57.

\noindent{} Dickey, J. M. \& Lockman, F. J. 1990, ARAA, 28, 215.

\noindent{} Eracleous, M., Halpern, J. \& Patterson, J. 1991, ApJ, 382, 290.

\noindent{} Downes, R. A., Webbink, R. F. \& Shara, M. M. 1997, PASP, 109, 345.

\noindent{} Hakala, P. et al. 2003, MNRAS, 343, 10L.

\noindent{} Israel, G. L. et al. 2002, A\&A, 386, L13.

\noindent{} Kondo, M., Noguchi, T. \& Maehara, H. 1984, Ann. Tokyo
Astron. Obs., 20, 130.

\noindent{} Marsh, T. R. \& Steeghs, D. 2002, MNRAS, 331, L7.

\noindent{} Marshall, H. L. et al. 2003, astro-ph/0308332.

\noindent{} Narayan, R. \& Popham, R, G. 1993, Nature, 362, 820.

\noindent{} Nelemans, G., Yungelson, L.~R., \& Portegies Zwart, S.~F.\
2004, MNRAS, 349, 181.

\noindent{} Norton, A. J., Haswell, C. A. \& Wynn, G. A. 2004, A\&A,
in press, (astro-ph/0206013).

\noindent{} Patterson, J. et al. 2002, PASP, 114, 65.

\noindent{} Popham, R. G. \& Narayan, R. 1995, ApJ, 442, 337.

\noindent{} Pringle, J. E. \& Savonije, G. J. 1979, MNRAS, 187, 777.

\noindent{} Ramsay, G. Hakala, P. \& Cropper, M. 2002, MNRAS, 332, L7.

\noindent{} Ramsay, G., Cropper, M., Wu, K., Mason, K.~O., \& Hakala, P.\ 
2000, MNRAS, 311, 75.

\noindent{} Strohmayer, T. E. 2004a, ApJ, in press, (astro-ph/0403675).

\noindent{} Strohmayer, T. E. 2004b, ApJ, in press, (astro-ph/0404542).

\noindent{} Strohmayer, T. E., 2003, ApJ, 593, 39L.

\noindent{} Strohmayer, T. E., 2002, ApJ, 581, 577.

\noindent{} Ulla, A. 1995, A\&A, 301, 469.

\noindent{} van Teeseling, A. \& Verbunt, F. 1994, A\&A, 292, 519.

\noindent{} Voges, W. et al. 2000,  
http://wave.xray.mpe.mpg.de/rosat/catalogues/rass-fsc/ 

\noindent{} Warner, B. 2003, to appear in the proceedings of IAU JD5,
    `White Dwarfs: Galactic and Cosmological Probes', eds. Ed Sion,
    Stephane Vennes and Harry Shipman, (astro-ph/0310243).

\noindent{} Warner, B. 1995, {\it Cataclysmic Variable Stars}, Cambridge
Univ. Press, Cambridge UK.

\noindent{} Warner, B. \& Woudt, P. A. 2002, PASP, 114, 129.

\noindent{} Wegner, G., McMahon, R. K. \& Boley, F. I. 1987, AJ, 94, 1271.

\noindent{} Woudt, P. A. \& Warner, B. 2003, to appear in the
    proceedings of IAU JD5, `White Dwarfs: Galactic and Cosmological
    Probes', eds. Ed Sion, Stephane Vennes and Harry Shipman,
    (astro-ph/0310494).

\noindent{} Wu, K., Cropper, M., Ramsay, G. \& Sekiguchi, K. 2002,
MNRAS, 331, 221.

\pagebreak

\begin{figure}
\begin{center}
 \includegraphics[width=5.0in, height=4.5in]{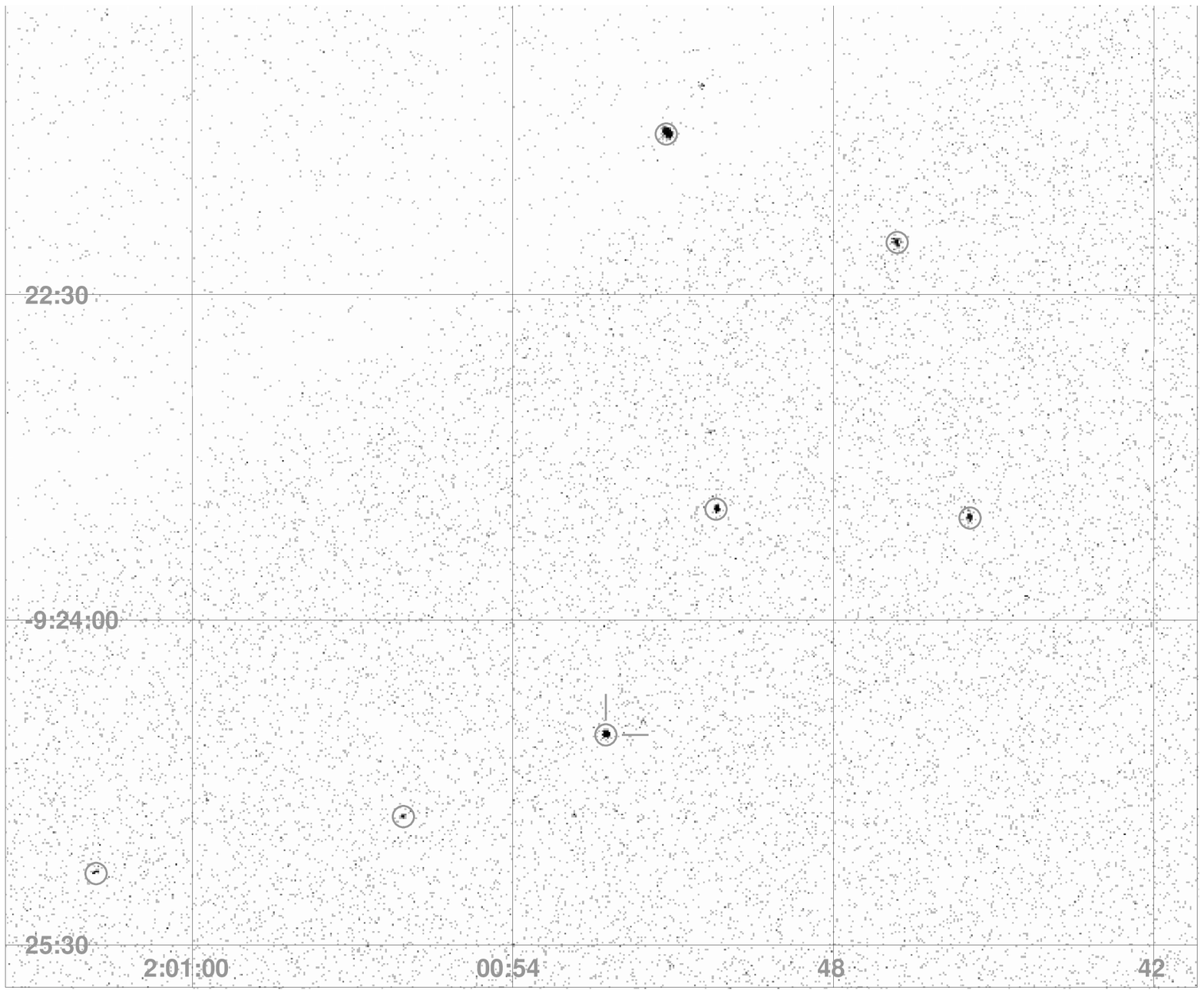}
\end{center}
Figure 1: Chandra/ACIS-S 0.2 - 10 keV image of the region around ES
Cet. Seven previously unknown sources are identified with 3'' radius
circles. The position of ES Cet is additionally marked by the pair of
orthogonal lines.
\end{figure}
\clearpage

\begin{figure}
\begin{center}
\includegraphics[width=6in, height=6in]{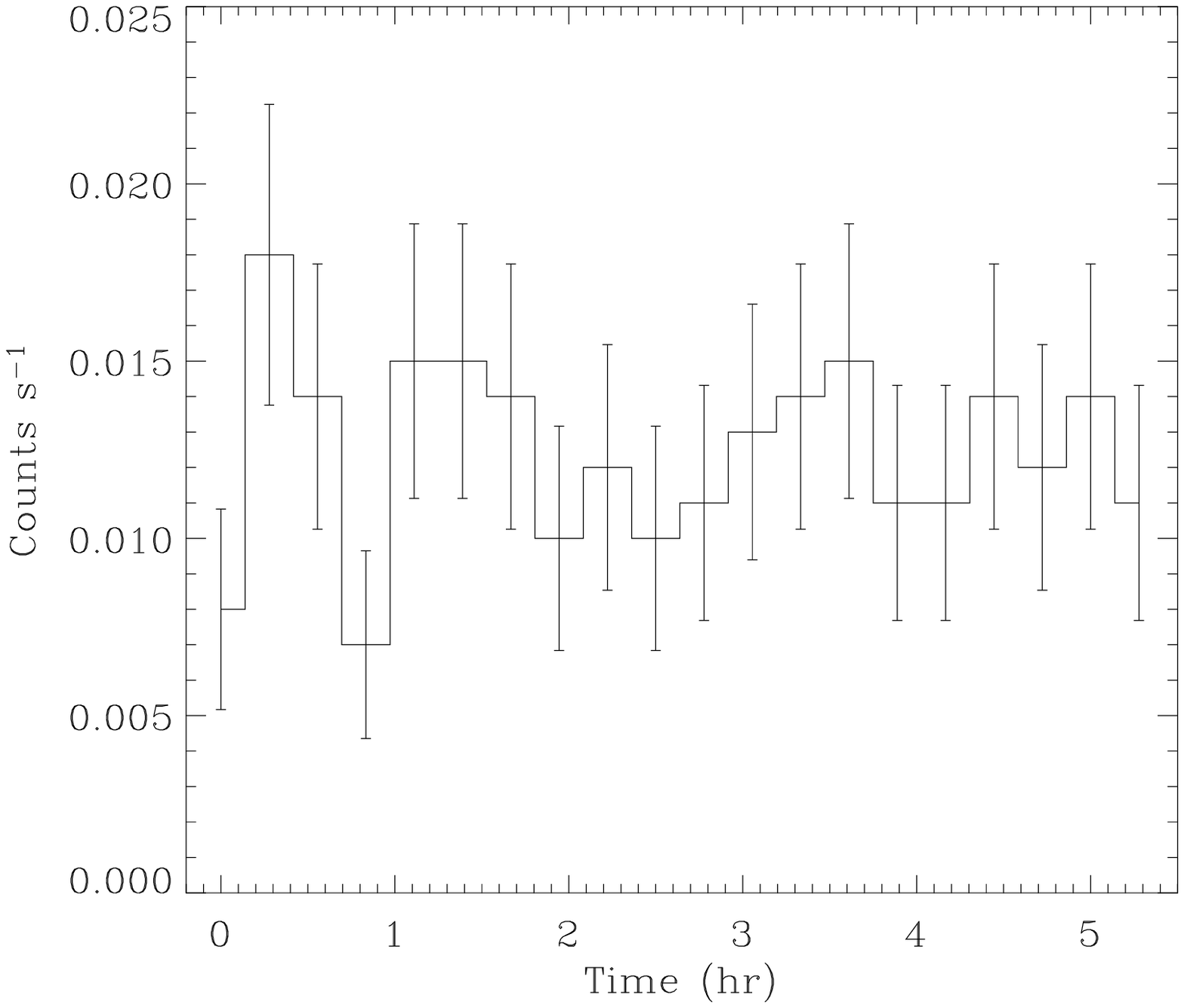}
\end{center}
Figure 2: Lightcurve of ES Cet in the full {\it Chandra} band. The
time bins are 1000 s. There is no evidence for variability.
\end{figure}
\clearpage

\begin{figure}
\begin{center}
 \includegraphics[width=6in, height=6in,angle=-90]{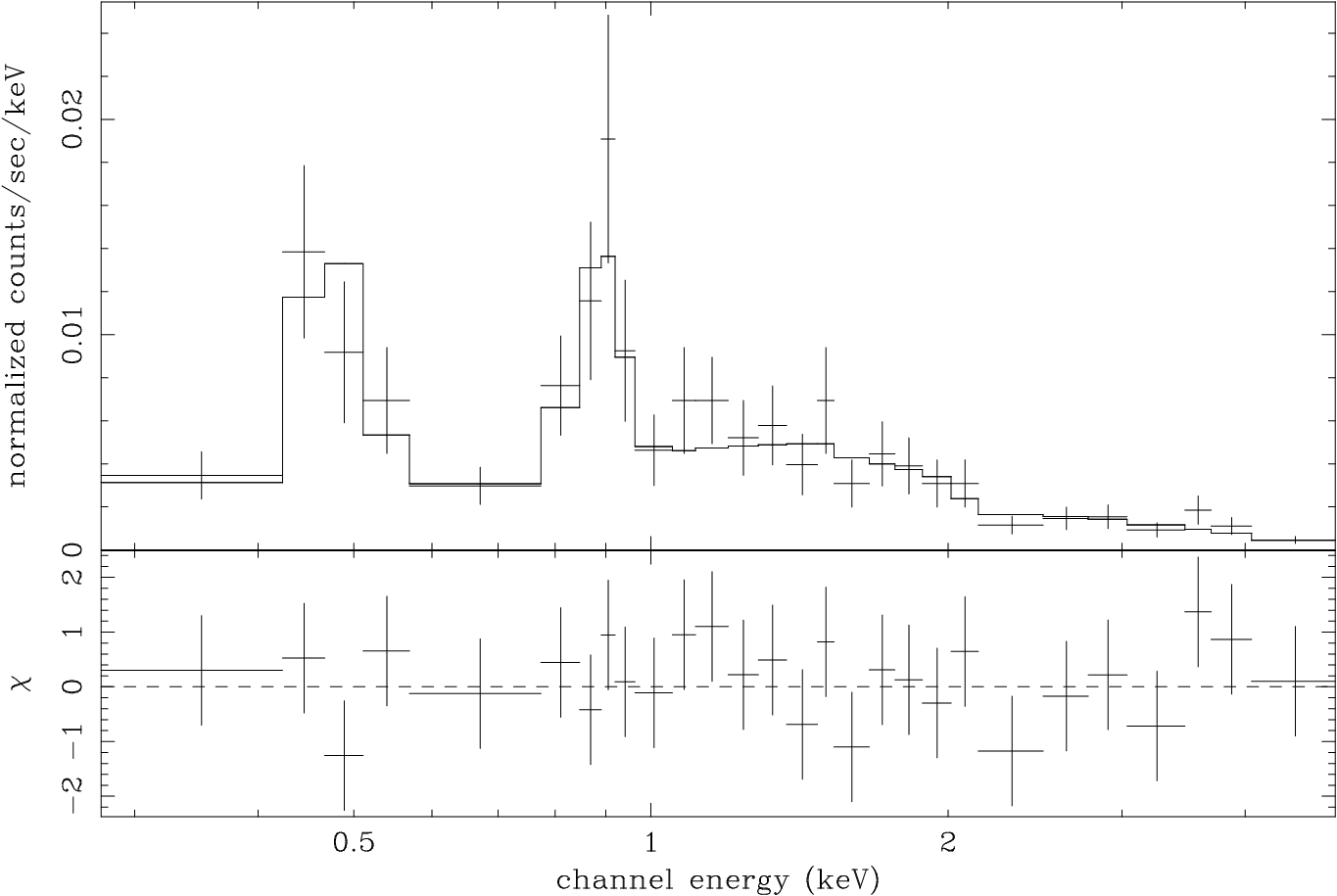}
\end{center}
Figure 3: ACIS-S data and best fitting spectral model. The model
includes black body emission with $kT = 0.8$ keV, as well as narrow
gaussian emission lines at 470 and 890 eV. See Table 2 for details of
the spectral modeling.
\end{figure}
\clearpage

\begin{table*}
\begin{center}{Table 1: X-ray Sources in the Vicinity of ES Cet}
\begin{tabular}{cccc} \\
\tableline
\tableline
 Source &  RA (J2000) & DEC (J2000) & Count Rate (s$^{-1}$)$^1$ \\
\tableline
 CXO 020052.2-092431.6 (ES Cet) & 02:00:52.23 & -09:24:31.68 & 0.0101 \\
 CXO 020056.0-092454.4 & 02:00:56.05 & -09:24:54.40 & 0.0011 \\
 CXO 020050.1-092329.2 & 02:00:50.18 & -09:23:29.26 & 0.0045 \\
 CXO 020101.8-092509.5 & 02:01:01.82 & -09:25:09.52 & 0.0012 \\
 CXO 020051.1-092145.3 & 02:00:51.10 & -09:21:45.32 & 0.0012 \\
 CXO 020046.8-092215.4 & 02:00:46.81 & -09:22:15.49 & 0.0007 \\
 CXO 020045.4-092331.6 & 02:00:45.43 & -09:23:31.60 & 0.0027 \\
\tableline
\end{tabular}
\end{center}
$^1$ Background subtracted count rate.
\end{table*}

\clearpage

\begin{table*}
\begin{center}{Table 2: X-ray Spectral Fits for ES Cet}
\begin{tabular}{ccccccccc} \\
\tableline 
\tableline 
Model & $kT/\alpha^1$ & Norm$^2$ & Flux$^3$ &
E$_1^4$ & Flux$_{E_1}^5$ & E$_2^4$ & Flux$_{E_2}^5$ & $\chi^2_r$ \\
\tableline 
powerlaw & $0.95 \pm 0.1$ & $1.1\times 10^{-5}$ & $6.7$ & -
& - & - & - & 1.4 \\ 
blackbody & $0.68 \pm 0.05$ & 0.029 & $6.1$ & - & - & - & - & 1.84 \\ 
powerlaw + lines & $ 0.88 \pm 0.14$ & $8.1\times
10^{-6}$ & $7.1$ & $471 \pm 29$ & $3.1$ & $891 \pm 29$ & $1.9$ & 0.95 \\ 
blackbody + lines & $ 0.79 \pm 0.08$ & 0.017 & $6.9$ & $470 \pm 19$
& $4.7$ & $890 \pm 23$ & $2.3$ & 0.60 \\ 
\tableline
\end{tabular}
\end{center}
$^1$ Temperature in keV (for blackbody), or power law index.

$^2$ Continuum normalization parameter.

$^3$ Model flux in units of $10^{-14}$ ergs cm$^{-2}$ s$^{-1}$.

$^4$ Line centroid in eV.

$^5$ Line flux in units of $10^{-6}$ photons cm$^{-2}$ s$^{-1}$.

\end{table*}


\end{document}